\documentclass{aa}  

\usepackage{graphicx}
\usepackage{txfonts}
\usepackage{epsfig}
\usepackage{longtable}
\usepackage{amssymb} 
\newcommand{\sw}{$Swift$}

\def \inte {\emph{INTEGRAL}}
\def \xmm {\emph{XMM-Newton}}

\def \sw {{\it Swift}}
\def \chandra {{\it Chandra}}

\def \src {\mbox{IGR~J17354--3255}}

\def \hcm {\hbox {\ifmmode $ atom cm$^{-2}\else atom cm$^{-2}$\fi}}
\def \arcmin {\hbox{$^\prime$}}
\def \arcsec {\hbox{$^{\prime\prime}$}}

\def \apjl {ApJL}
\def \apjs {ApJS}

\def \aap {A\&A}

\begin{document}

   \title{\emph{Swift}/XRT orbital monitoring of the candidate 
supergiant fast X--ray transient IGR J17354--3255}


   \author{L. Ducci
          \inst{1}
          \and
          P. Romano\inst{2}
          \and
          P. Esposito\inst{3}
          \and
          E. Bozzo\inst{4}
          \and
          H.A. Krimm\inst{5,6}
          \and
          S. Vercellone\inst{2}
          \and
          V. Mangano\inst{2}
          \and
          J.A.~Kennea\inst{7}
          }
   \institute{Institut f\"ur Astronomie und Astrophysik, Eberhard Karls Universit\"at, 
              Sand 1, 72076 T\"ubingen, Germany\\
              \email{ducci@astro.uni-tuebingen.de}
              \and
              INAF, Istituto di Astrofisica Spaziale e Fisica Cosmica - Palermo,
              Via U.\ La Malfa 153, I-90146 Palermo, Italy
              \and
              INAF, Istituto di Astrofisica Spaziale e Fisica Cosmica - Milano, 
              Via E.\ Bassini 15,   I-20133 Milano,  Italy 
              \and
              ISDC, Data Center for Astrophysics of the University of Geneva, 
              Chemin d'Ecogia, 16 1290 Versoix Switzerland
              \and
              NASA/Goddard Space Flight Center, Greenbelt, MD 20771, USA
              \and
              Universities Space Research Association, Columbia, MD, USA 
              \and
              Department of Astronomy and Astrophysics, Pennsylvania State 
              University, University Park, PA 16802, USA
             }

   \date{}

 
  \abstract
{We report on the \sw/X-ray Telescope (XRT) monitoring 
of the field of view around the candidate supergiant 
fast X-ray transient (SFXT) IGR~J17354--3255, which is positionally associated with the 
\emph{AGILE}/GRID gamma-ray transient AGL~J1734--3310. 
Our observations, which cover 11 days for a total on-source exposure of $\sim$  24 ks, 
span 1.2 orbital periods ($P_{\rm orb}=8.4474$ d) and are the first sensitive monitoring 
of this source in the soft X--rays.  
These new data allow us to exploit the timing variability properties 
of the sources in the field to unambiguously identify 
the soft X--ray counterpart of IGR~J17354--3255.
The soft X-ray light curve shows a moderate orbital modulation and 
a dip.
We investigated the nature of the dip by comparing the X-ray light curve with the prediction
of the Bondi-Hoyle-Lyttleton accretion theory,
assuming both spherical and nonspherical symmetry of the outflow from the donor star. 
We found that the dip cannot be explained with the X-ray orbital modulation.
We propose that an eclipse or the onset of a gated mechanism is the most likely explanation
for the observed light curve.}


   \maketitle
%

	\section{Introduction\label{igr17354:intro}}

\src\ was discovered as a hard X-ray transient on 2006 April 21  
\citep[][]{Kuulkers2006:atel874}, when it reached a flux of about 18 mCrab ($20-60$ keV)
during two $\sim 1.8$ ks observations within the \inte\ Galactic bulge monitoring program. 
It  was first reported in the fourth IBIS/ISGRI catalog \citep{Bird2010:igr4cat_mn} 
with a 20--40 keV average flux of $1.4\times10^{-11}$ erg cm$^{-2}$ s$^{-1}$.  
It is listed in the \sw\ \citep[][]{Gehrels2004mn} Burst Alert Telescope \citep[BAT, ][]{Barthelmy2005:BAT} 
58-month Hard X-ray Survey \citep[][Swift J1735.6--3255]{Baumgartner2010:BAT58mos}
with a 14--195 keV average flux of $2.7\times10^{-11}$ erg cm$^{-2}$ s$^{-1}$
and in the 54-month Palermo \sw\ BAT hard X-ray catalog 
\citep[][2PBC~J1735.4--3256]{Cusumano2010:batsur_III} 
with a 15--150 keV average flux of $(2.12\pm1.15)\times10^{-11}$ erg cm$^{-2}$ s$^{-1}$.  

Renewed interest in this object arose due to its positional association with 
the \emph{AGILE}/GRID gamma-ray transient AGL~J1734--3310 \citep[][]{Bulgarelli2009:atel2017mn}. 
Thanks to a follow-up \sw\ target of opportunity (ToO) 
observation of the \src\ field on 2009 April 17, 
\citet[][]{Vercellone2009:atel2019mn} detected two sources 
(hereafter src1 and src2) within the 
4$\arcmin$ \inte\ error circle \citep[][]{Kuulkers2006:atel874}  
at a 0.2--10 keV count rate of $\sim 0.1$ counts s$^{-1}$  (src1) 
and $\sim 0.005$ counts s$^{-1}$ (src2). 
Based on a comparison with a previous observation of the same field 
taken on 2008 March 11 in which 
src1 was not detected while src2 showed the same count rate,
\citet[][]{Vercellone2009:atel2019mn} suggested that src2 
was a persistent source and src1 the most likely counterpart of \src. 
\chandra\ observations \citep[][]{Tomsick2009:cxc17354} also found two sources within the
\inte\ error circle. CXOU~J173527.5-325554 (src1) was the brighter 
(unabsorbed 0.3--10 keV flux 
of $\sim1.3\times10^{-11}$ erg cm$^{-2}$ s$^{-1}$), harder ($\Gamma\sim0.6$), and more heavily 
absorbed ($N_{\rm H}=7.5\times10^{22}$ cm$^{-2}$, $N_{\rm Gal}=1.2\times10^{22}$ cm$^{-2}$), 
arguing for a high-mass X-ray binary (HMXB) nature.  
With an unabsorbed 0.3--10 keV flux of $\sim1.4\times10^{-12}$ erg cm$^{-2}$ s$^{-1}$,
CXOU~J173518.7-325428 (src2) could also contribute to the flux detected by \inte,
but \citet[][]{Tomsick2009:cxc17354} also suggested that CXOU~J173527.5-325554 (src1) 
was the counterpart of \src. 

\citet[]{Dai2011:period_17354} and \citet[][]{Sguera2011:17354} have 
studied the hard X--ray properties of \src\ as they emerge from the extensive BAT 
and \inte\ archives, respectively.
They thus discovered a modulation of the light curve with a period of
$8.4474\pm0.0017$ d \citep[][]{Sguera2011:17354}, which they interpret as the 
orbital period of the binary. In both works it is shown that the 2009 
\sw/X-ray Telescope \citep[XRT, ][]{Burrows2005:XRTmn} observation was 
performed at the maximum of the folded light curve, 
while the 2008 one was performed at the minimum,
thus strengthening the association of src1 with \src. 
Because of its hard X-ray flaring activity 
(mean flare flux of 20--40 mCrab with one flare peaking at 
108 mCrab in the 18--60 keV band) and large dynamic range, 
\citet[][]{Sguera2011:17354} proposed \src\ as a candidate 
supergiant fast X-ray transient (SFXT) and 
investigate its association with the transient AGL~J1734--3310. 

The SFXTs are HMXBs firmly associated 
with O or B supergiant stars through optical spectroscopy
\citep[e.g., ][]{Negueruela2006:ESASP604}, which display short 
\citep[just a few hours long, as observed by \inte; ][]{Sguera2005,Negueruela2006}  
X--ray outbursts with peak luminosities of 10$^{36}$--10$^{37}$~erg~s$^{-1}$ 
and an overall dynamic range of 3--5 orders of magnitude. 
Their hard X--ray spectra during outburst resemble those of HMXBs 
hosting accreting neutron stars, with often heavily absorbed, 
hard power laws below 10 keV combined 
with high-energy cut-offs at $\sim 15$--30~keV. 
While the detailed mechanism producing the outbursts is not well established,  
it is probably related to either the properties of 
the wind from the supergiant companion 
\citep{zand2005,Walter2007,Negueruela2008,Sidoli2007} or the 
presence of a centrifugal or magnetic barrier 
\citep[][]{Grebenev2007,Bozzo2008}. 
Recently, some evidence has been accumulating that
indicates that SFXTs may be the X-ray counterparts of a new class of 
short Galactic transients emitting in the MeV/GeV regime \citep[see, e.g.,][]{Sguera2009}.

In this paper we analyze all the \sw/XRT data collected on the region of \src\ (Sect. \ref{igr17354:dataredu}). 
This allows us to unambiguously identify the soft X--ray counterpart of the candidate SFXT \src\ 
with src1, based on its timing properties (Sect. \ref{igr17354:results}).  
Furthermore, we show the results of the most intense and complete sampling along the orbital 
period of the light curve of this candidate SFXT with a sensitive soft X-ray instrument 
and discuss the nature of the companion (Sect. \ref{igr17354:discuss}).

 \begin{table}
 \begin{center}
\tabcolsep 3pt   
 \caption{Summary of the {\it Swift}/XRT observations.\label{igr17354:tab:xrtobs} }
\resizebox{\columnwidth}{!}{
\begin{tabular}{lllll}
 \hline
 \hline
 \noalign{\smallskip}
 ObsID     & Start time  (UT)  & End time   (UT) & Exp. & $\phi^a$   \\ 
              &   &  &(s)        \\
  \noalign{\smallskip}
 \hline
 \noalign{\smallskip}
00037054001  &2008-03-11 01:05:20     &      2008-03-11 10:52:57    &      4403          &  0.59  \\  
00037054002  &2009-04-17 01:08:59     &      2009-04-17 07:57:56    &      5243          &  0.17  \\
00032513001	&	2012-07-18 04:36:12	&	2012-07-18 04:52:55	&	980	           &  0.80 \\   
00032513002	&	2012-07-18 11:00:16	&	2012-07-18 11:16:54	&	978	           &  0.83  \\  
00032513003	&	2012-07-18 15:58:29	&	2012-07-18 16:18:56	&	1216	   &  0.86  \\ 
00032513004	&	2012-07-18 20:56:48	&	2012-07-18 21:13:55	&	1013	   &  0.89  \\ 
00032513006	&	2012-07-19 06:10:32	&	2012-07-19 06:30:56	&	1209	   &  0.93  \\ 
00032513007	&	2012-07-19 17:25:34	&	2012-07-19 17:35:56	&	622	           &  0.99  \\ 
00032513008	&	2012-07-20 11:02:19	&	2012-07-20 15:59:55	&	1010	   &  0.08  \\ 
00032513009	&	2012-07-21 15:57:24	&	2012-07-21 23:59:54	&	1078	   &  0.23 \\
00032513012	&	2012-07-23 01:34:36	&	2012-07-23 03:17:54	&	1103	   &  0.38  \\
00032513011	&	2012-07-23 08:03:03	&	2012-07-23 08:21:54	&	1126	   &  0.41  \\ 
00032513010	&	2012-07-23 14:24:36	&	2012-07-23 16:10:54	&	880	           &  0.45  \\ 
00032513014	&	2012-07-24 04:52:54	&	2012-07-24 04:58:27	&	311	           &  0.52  \\ 
00032513016	&	2012-07-24 14:35:14	&	2012-07-24 14:58:56	&	1402	   &  0.56  \\ 
00032513017	&	2012-07-24 21:15:56	&	2012-07-24 22:59:55	&	1063	   &  0.60  \\ 
00032513018	&	2012-07-25 01:40:49	&	2012-07-25 02:20:55	&	1013	   &  0.62  \\ 
00032513019	&	2012-07-25 08:25:10	&	2012-07-25 08:45:56	&	1226	   &  0.65  \\ 
00032513020	&	2012-07-25 14:36:48	&	2012-07-25 14:57:54	&	1266	   &  0.68  \\ 
00032513021	&	2012-07-25 21:19:00	&	2012-07-25 23:04:54	&	1289	   &  0.72  \\ 
00032513022	&	2012-07-27 03:23:09	&	2012-07-27 03:44:55	&	1284	   &  0.86  \\ 
00032513023	&	2012-07-27 09:47:09	&	2012-07-27 10:08:56	&	1286	   &  0.90  \\ 
00032513024	&	2012-07-27 13:00:09	&	2012-07-27 13:19:55	&	1171	   &  0.91  \\ 
00032513025	&	2012-07-28 01:50:20	&	2012-07-28 02:29:54	&	1191	   &  0.98  \\ 
  \noalign{\smallskip} 
  \hline
  \end{tabular}
}
  \end{center}
  $^{\mathrm{a}}$ Mean phase referred to  $P_{\rm orb}=8.4474$ d and $T_{\rm epoch}=$ MJD 52698.205.
\end{table}

\begin{figure}
\begin{center}
\vspace{-0.5truecm}
\includegraphics*[angle=0,width=\columnwidth]{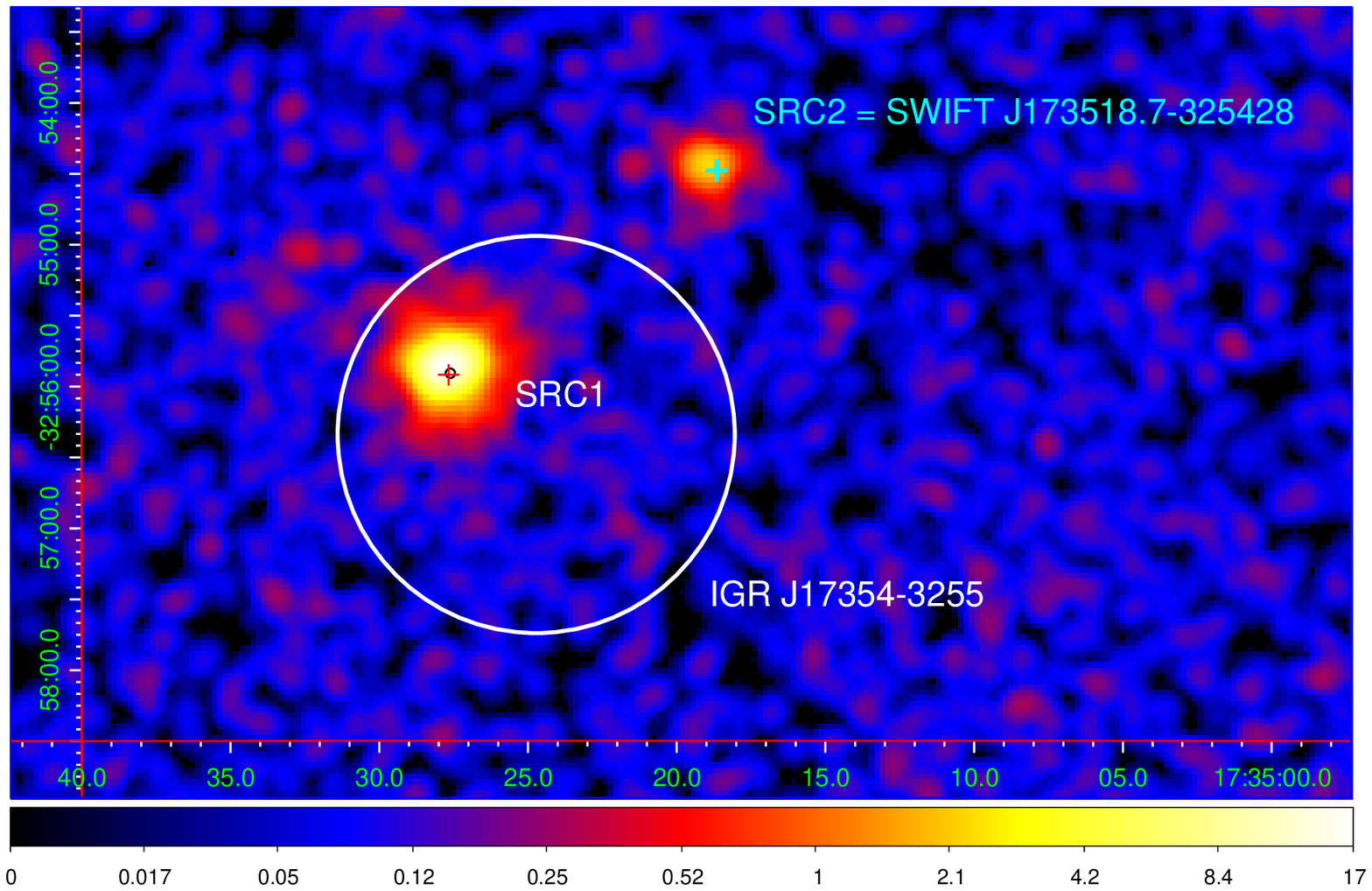}
\end{center}
\vspace{-0.3truecm}
\caption{Region around \src, as observed by \sw/XRT (0.3--10 keV). All data collected by XRT were used.
The large circle is the \inte\ 90\% error box \citep[1.4$\arcmin$,][]{Bird2010:igr4cat_mn}.
The crosses mark the two XRT sources in the field of \src\ (src1 within the \inte\ error circle, src2 outside of it).
The black circle represents the position of the 2MASS counterpart of \src. 
}
\label{igr17354:fig:map}
\end{figure}

\begin{figure}
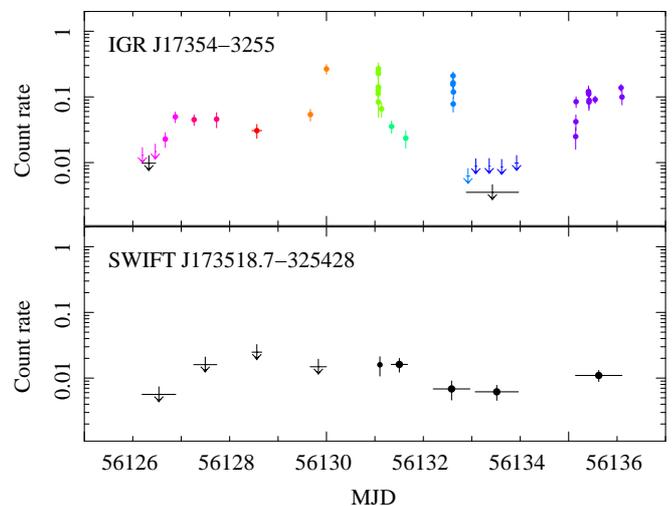

\begin{center}
\includegraphics[angle=270,width=9cm]{figure2a.ps}

\vspace{-0.05truecm} 
\includegraphics[angle=270,width=9cm]{figure2b.ps}
\end{center}
\caption{\sw/XRT 0.3--10 keV light curves of the two X-ray sources detected 
within the region around \src,  during the 2012 July monitoring program. 
Downward-pointing arrows are 3$\sigma$ upper limits. 
Top: Swift~J173527.7--325555 (src1, identified with the soft X-ray counterpart of \src).  
Different colors 
mark different days 
(see Table~\ref{igr17354:tab:xrtobs}). Bottom: Swift~J173518.7--325428 (src2). 
}
\label{igr17354:fig:xrtlcv_mjd}
\end{figure}

 	 \section{Observations and Data Analysis \label{igr17354:dataredu}}

We first considered the hard X-ray data from the \sw/BAT. 
\src\ has never triggered BAT onboard, but it is detected in the Hard X-ray 
Survey\footnote{http://swift.gsfc.nasa.gov/docs/swift/results/bs58mon/SWIFT\_J1735.6-3255} 
at 9.6$\sigma$ confidence level. The light curve, which spans the 65 months between 
MJD 53373.66 and 55347.65 (2005 January 3 to 2010 May 31), 
shows a fairly steady level with no significant long-term variability. 
Two short flares are found, with fluxes consistent with \inte-detected flares, on 
2006 March 4 22:44 UT (5.6$\sigma$, lasting 1044 s),  
when the source reached 0.017 counts s$^{-1}$ cm$^{-2}$ (14--195 keV), and on 
2006 March 25 18:29 UT (5.7$\sigma$, 1324 s), reaching  0.010 counts s$^{-1}$ cm$^{-2}$. 

We also retrieved the IBIS/ISGRI light curve of \src\ from the online tool 
Heavens\footnote{http://www.isdc.unige.ch/heavens}, 
which has access to all public available \inte\ data on this source,
between MJD 52671.43 and 55624.28 (2003 February 1 to
2011 March 4). We selected the 17--50 keV energy band to maximize the
signal-to-noise ratio.

\begin{figure*}
\begin{center}
\includegraphics*[angle=270,width=17.5cm]{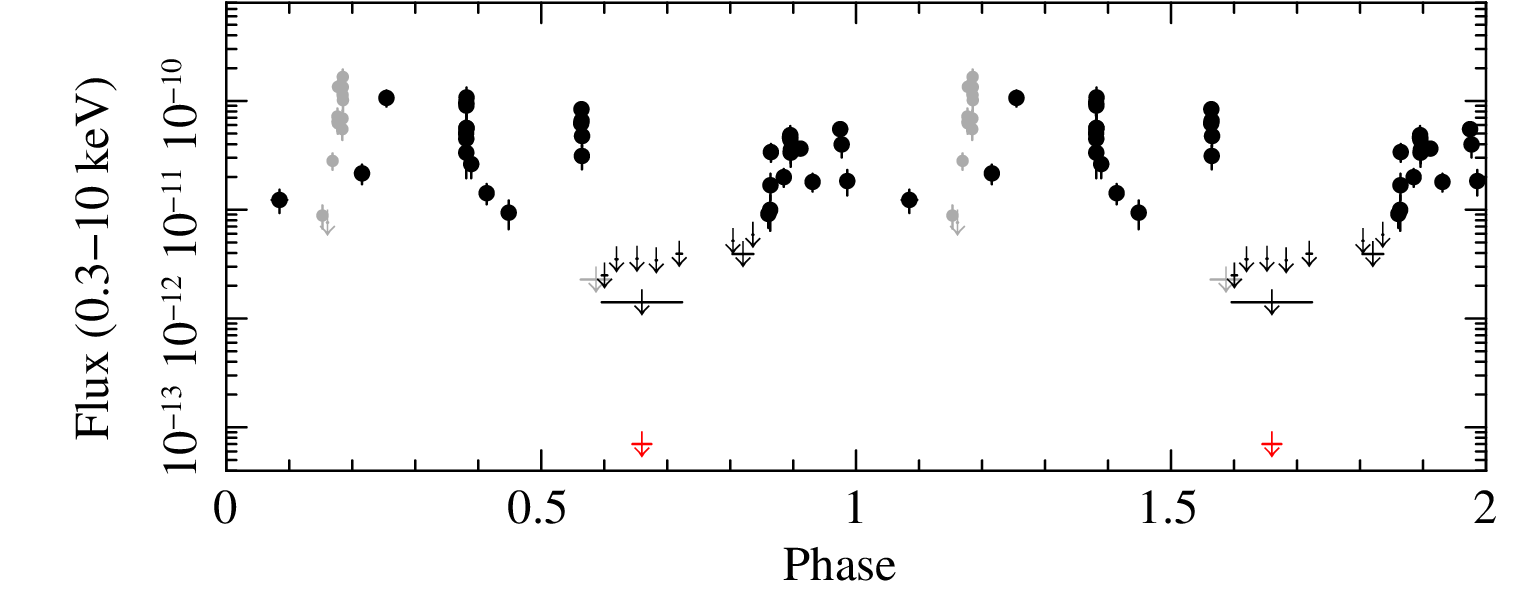}
\end{center}
\caption{\sw/XRT 0.3--10 keV flux light curve of \src, 
folded at $P_{\rm orb}=8.4474$ d and $T_{\rm epoch}=$ MJD 52698.205 \citep[][]{Sguera2011:17354}. 
Downward-pointing arrows are 3$\sigma$ upper limits. 
The data collected in 2012 are shown in black, the 2008 and 2009 data
in grey.
The large black downward-pointing arrow at phase 0.66 is obtained by
combining five observations (00032513017 through 00032513021)
for a total on-source exposure of 5.9 ks. 
The black downward-pointing arrow at phase 0.82 is obtained by
combining two observations (00032513001--2) 
for a total on-source exposure of 2.0 ks. 
The red arrow 
at $\sim 7\times10^{-14}$ erg cm$^{-2}$ s$^{-1}$, also at phase 0.66,
is the  3-$\sigma$ upper limit obtained from a 19 ks exposure on 2011 March 6 
with  \xmm\ \citep[][]{Bozzo2012:HMXBs}.  
}
\label{igr17354:fig:xrtlcv_phase}
\end{figure*}

The log of the \sw/XRT 
observations used in this paper is in Table~\ref{igr17354:tab:xrtobs}. 
The \sw\ observations of \src\ during 2008 and 2009 were extensively described in 
\citet[][]{Vercellone2009:atel2019mn},  \citet[][]{Dai2011:period_17354}, and 
\citet[][]{Sguera2011:17354}; here they are re-analyzed to ensure uniformity. 
The data collected in 2012 July were obtained as a ToO 
monitoring program of ten scheduled daily observations, 
which were each 5 ks long and equally spread 
in the four quarters of the day, starting on  2012 July 18. 
This strategy aimed at covering the light curve in all 
orbital phases while keeping the observing time per day 
reasonably short, thus not impeding observations of gamma-ray bursts (GRBs), 
which are the main scientific target for \sw. 
The 2012 campaign lasted 11 days divided in 22 observations  
for a total on-source exposure of $\sim$  24  ks.

The XRT data were uniformly processed with standard procedures 
({\sc xrtpipeline} v0.12.6), as well as filtering and screening criteria by using 
{\sc FTOOLS v6.12}.  
Within the XRT field of view (FOV) around \src, two sources were detected
 (see Fig.~\ref{igr17354:fig:map}). 
In both cases the source count rates never exceeded 
$\sim0.5$ count s$^{-1}$, so only photon-counting mode (PC) events (selected in grades 0--12) 
were considered and we checked that no pile-up correction was required. 
Source events were accumulated within a circular region 
with a radius of 20 pixels (1 pixel $\sim2.36$\arcsec),  
and background events were accumulated from a nearby source-free region.  

Light curves were created for both sources and corrected 
for point-spread function (PSF) losses, 
vignetting, and were background subtracted (see Fig.~\ref{igr17354:fig:xrtlcv_mjd}).
For src1 the light curve was binned to ensure at least 20 counts per 
bin, whenever the statistics allowed it. 
For src2 we accumulated all data within a day to obtain detections whenever possible. 
For our spectral analysis, we extracted events in the same regions as 
those adopted for the light curve creation; ancillary response files were 
generated with {\sc xrtmkarf}.  
We used the latest spectral redistribution matrices in CALDB (20120713). 
For a more detailed discussion of the data analysis procedure, please refer to 
\citet[][and references therein]{Romano2011:sfxts_paperVI}.

All quoted uncertainties are given at 90\% confidence level (c.l.) for 
one interesting parameter, unless otherwise stated.

 	 \section{Results \label{igr17354:results} }

Swift~J173527.7--325555 (src1) is located at 
RA(J$2000)=17^{\rm h}\,  35^{\rm m}\,  27\fs66$, 
Dec(J$2000)=-32^{\circ}\,  55^{\prime}\, 55\farcs1$ 
with an uncertainty radius of 2\farcs0  (90\% c.l.;
the astrometrically corrected position was determined according to 
\citealt[][]{Evans2009:xrtgrb_mn} and \citealt[][]{Goad2007:xrtuvotpostions}). 
This position is $1\farcs1$ from the \chandra\  unidentified source 
CXOU~J173527.5--325554 \citep{Tomsick2009:cxc17354} and 
$0\farcs9$ from the source 2MASS ~J17352760--3255544.  
In the following we report on src1, while the properties of
src2 are described in Sect.~\ref{igr17354:src2}.

 	 \subsection{Light curve analysis  \label{igr17354:timing}}

\begin{figure}
\begin{center}
\hspace{-0.8truecm}
\includegraphics*[angle=270,width=9cm]{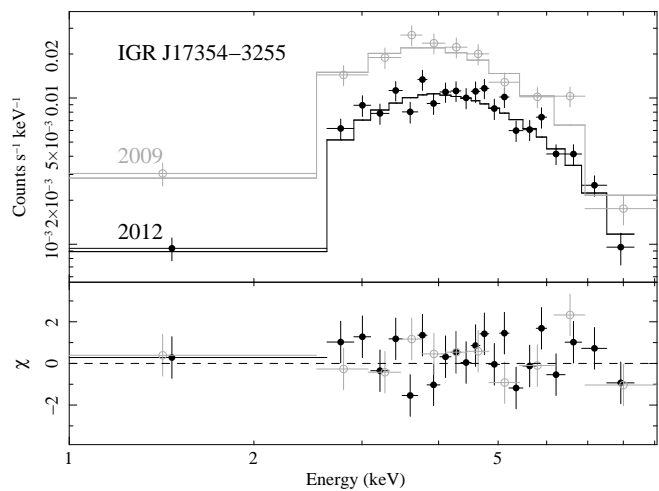}
\end{center}
\caption{Spectrum of the 2009 observation (grey) and 2012 campaign (black).    
                Top:  data fit with an absorbed power-law model (see Table~\ref{igr17354:tab:xrtspec}). 
                Bottom: the residuals of the fits (in units of standard deviations). 
}
\label{igr17354:fig:spec}
\end{figure}

Fig.~\ref{igr17354:fig:xrtlcv_mjd} (top) shows the 
0.3--10 keV light curve of Swift~J173527.7--325555 during  the 2012 campaign. 
The light curve begins at phase 0.8 (see Table \ref{igr17354:tab:xrtobs})
and covers the source through a little over one full period, 
ending at phase 0.98.  
Fig.~\ref{igr17354:fig:xrtlcv_phase} shows the  folded light curve,
which includes the data collected in 2012 (black) and the 2008 and 2009 data (grey). 
We also include the  3-$\sigma$ upper limit obtained from a 19 ks exposure 
on 2011 March 6 with  \xmm\ \citep[][]{Bozzo2012:HMXBs}.  
The XRT count rate to flux conversion was derived from the spectral fit of the mean 2012
spectrum (see Sec.~\ref{igr17354:spectroscopy}). 
Superimposed on the long-term orbital modulation, 
which follows the one seen in the BAT \citep[][fig.~3c]{Dai2011:period_17354} and 
\inte\ data \citep[][fig.~4]{Sguera2011:17354} we also observe  
flaring on short time scales (hundred of seconds, e.g., grey points in Fig.~\ref{igr17354:fig:xrtlcv_phase}). 
This behavior has been observed in most SFXTs  
\citep[][]{Sidoli2008:sfxts_paperI,Romano2009:sfxts_paperV,Romano2011:sfxts_paperVI},
in particular those we monitored along the orbital period 
\citep[][]{Romano2010:sfxts_18483,Romano2012:sfxts_16418}.

A noteworthy feature is the dip centered at $\phi \sim
0.7$, which starts at phase $\phi \sim 0.60$ 
and lasts $\Delta \phi \sim 0.2$--$0.24$. 
The lowest limit (as observed by XRT) was collected during the dip 
by combining five observations (00032513017 through 00032513021
for a total on-source exposure of 5.9 ks, shown as a larger
downward-pointing arrow at $\phi\sim0.66$). 
These observations yielded a 3-$\sigma$ upper limit at  $3.5\times10^{-3}$ counts s$^{-1}$. 
This corresponds to an unabsorbed 0.3--10 keV flux 
of $\sim 1.4 \times10^{-12}$ erg cm$^{-2}$ s$^{-1}$ 
and to a 2--10 keV luminosity of $8\times10^{33}$ erg s$^{-1}$ 
(assuming the optical counterpart distance of 8.5 kpc, as derived from the closeness 
to the Galactic center and the high X-ray absorption; \citealt[][]{Tomsick2009:cxc17354}).  
In addition, we note that the recent \xmm\ observation \citep[][]{Bozzo2012:HMXBs},
which was also obtained at phase 0.66, yielded a 3-$\sigma$ upper limit at 
$2\times10^{-3}$ counts s$^{-1}$ (0.5--10 keV)  
for the Metal Oxide Semi-conductor (MOS) cameras. 
This corresponds to an unabsorbed 0.3--10 keV flux of 
$\sim 7\times10^{-14}$ erg cm$^{-2}$ s$^{-1}$ when adopting the best-fitting spectral model   
for the 2012 XRT campaign (see Sect.~\ref{igr17354:spectroscopy}).  
The \xmm\ limit corresponds to a 2--10 keV luminosity of $4\times10^{32}$ erg s$^{-1}$. 

The lowest point during the campaign {\it outside} the dip was recorded on 
MJD 56126.67 at 0.023 counts s$^{-1}$, a detection corresponding to 
an unabsorbed 0.3--10 keV flux of $9.1\times10^{-12}$ erg cm$^{-2}$ s$^{-1}$
and to a luminosity of $5\times10^{34}$ erg s$^{-1}$. 
During the 2012 campaign, the peak count rate was $\ga 0.3$ counts s$^{-1}$ (MJD 56130.00).
The highest recorded count rate, however, 
was achieved on 54938.33 at $\ga 0.42$ counts s$^{-1}$
or $\sim 1.7 \times10^{-10}$ erg cm$^{-2}$ s$^{-1}$ ($9\times10^{35}$ erg s$^{-1}$). 

After having converted the event arrival times to the Solar system
barycentric frame, we searched the XRT data for periodic pulsations 
longer than $\sim 5$~s by means of a Fourier transform. 
No significant signal was found. 
Because of the limited statistics and the presence of substantial noise in the
time series, the upper limits on the source pulsed fraction are
non constraining (being larger than 100\%).

The observed orbital modulation in the 
folded light curves of this XRT source and \src,
which has a dip corresponding to the minimum observed in the folded 
BAT and \inte\ light curves (Fig. \ref{igr17354:fig:folded}),
allows a definitive identification of Swift~J173527.7--325555 as 
the soft X--ray counterpart of \src. This identification was  
previously mainly based on positional association.

 	 \subsection{Spectroscopy     \label{igr17354:spectroscopy}  }

  \begin{table} 	
\tabcolsep 3pt   
 \begin{center} 	
 \caption{XRT spectroscopy of \src. \label{igr17354:tab:xrtspec}} 	
 \label{} 	
 \begin{tabular}{lcccrr} 
 \hline
 \hline 
 \noalign{\smallskip} 
 ObsID      &$N_{\rm H}$  &$\Gamma$  &Flux$^a$     &L$^{b}$ &$\chi^{2}_{\nu}/$dof    \\
\noalign{\smallskip} 
           & (10$^{22}$~cm$^{-2}$) &           &            &  &   \\
 \hline 
\smallskip
00037054002             &$7.80_{-2.16}^{+2.91}$  &$1.65_{-0.53}^{+0.63}$ &$2.0$ & 1.7 &$1.04/18$  \\
\smallskip
00032513001--25     &$10.7_{-2.1}^{+2.6}$     &$1.68_{-0.44}^{+0.48}$  &$1.0$& 0.9 &$1.17/42$   \\
\noalign{\smallskip} 
  \hline
  \end{tabular}
  \end{center}
  \begin{list}{}{} 
  \item[$^{\mathrm{a}}$ Unabsorbed 2--10 keV fluxes ($10^{-11}$ erg~cm$^{-2}$~s$^{-1}$). ] 
  \item[$^{\mathrm{b}}$ 2--10 keV luminosities in units of $10^{35}$ erg~s$^{-1}$, at 8.5~kpc.] 
  \end{list} 
  \end{table} 

We extracted the mean spectrum during the 2009 observation (the grey points at 
phase 0.17 in Fig.~\ref{igr17354:fig:xrtlcv_phase}) 
and the mean spectrum of the 2012 campaign. 
The data were rebinned with a minimum of 20 counts per energy bin 
to allow $\chi^2$ fitting.  
The spectra were fit in the 0.3--10 keV energy range with a single absorbed 
power-law model as more complex models were not required by the data.  
The absorbing column is significantly in excess of the Galactic one
\citep[$1.59\times 10^{22}$ cm$^{-2}$; ][]{LABS}, while the photon index 
is $\Gamma\sim1.7$. The peak flux was reached in 2009 with its 
average 2--10 keV unabsorbed flux of $\sim 2\times 10^{-11}$ erg cm$^{-2}$ s$^{-1}$,
corresponding to $\sim 1.7\times10^{35}$  erg s$^{-1}$ at 8.5 kpc).
This is about a factor of 2 higher than the average spectrum obtained in 2012. 
The results are reported in Table~\ref{igr17354:tab:xrtspec} and the best fits are shown in 
Fig.~\ref{igr17354:fig:spec}.

 	 \subsection{Properties of Swift~J173518.7--325428    \label{igr17354:src2} }

Swift~J173518.7--325428 (src2) is located at 
RA(J$2000)=17^{\rm h}\, 35^{\rm m}\, 18\fs73$, 
Dec(J$2000)=-32^{\circ}\, 54^{\prime}\, 27\farcs5$, 
with an uncertainty radius of 4\farcs6  (90\% c.l.).  
This source is clearly outside the error circle of \src\
and is thus not a viable counterpart. 
For completeness, we report the analysis of this source.  
Our observations highlight a variability in its light curve, 
shown in Fig.~\ref{igr17354:fig:xrtlcv_mjd} (bottom), with a
dynamical range of 4.5 (at a binning of $\sim 1000$ s). 
The mean spectrum was extracted from all the data collected in 2012 and fit with an absorbed
power-law model in the 0.3--10 keV energy range with 
\citet{Cash1979} statistics. The fit yielded an absorbing column 
$N_{\rm H}=1.33_{-0.65}^{+0.84}\times 10^{22}$ cm$^{-2}$, 
which is consistent with the Galactic one \citep[$1.59\times 10^{22}$ cm$^{-2}$; ][]{LABS},
and  $\Gamma=1.1_{-0.5}^{+0.6}$. 
The average observed and unabsorbed fluxes (0.3--10 keV)
are $F_{\rm 0.3-10 keV}^{\rm obs}=7.1\times10^{-13}$ erg cm$^{-2}$ s$^{-1}$ 
and $F_{\rm 0.3-10 keV}=9.2\times10^{-13}$ erg cm$^{-2}$ s$^{-1}$, respectively, 
and are consistent with the findings of \citet[][]{Tomsick2009:cxc17354}. 
Although the optical counterpart of Swift~J173518.7--325428 is not detected,
the observed properties of this source, 
as also noted by \citet[][]{Tomsick2009:cxc17354}, 
are consistent with an X--ray binary.

\begin{figure}
\begin{center}
\hspace{-1truecm}
\includegraphics*[angle=-90.,width=9.3cm]{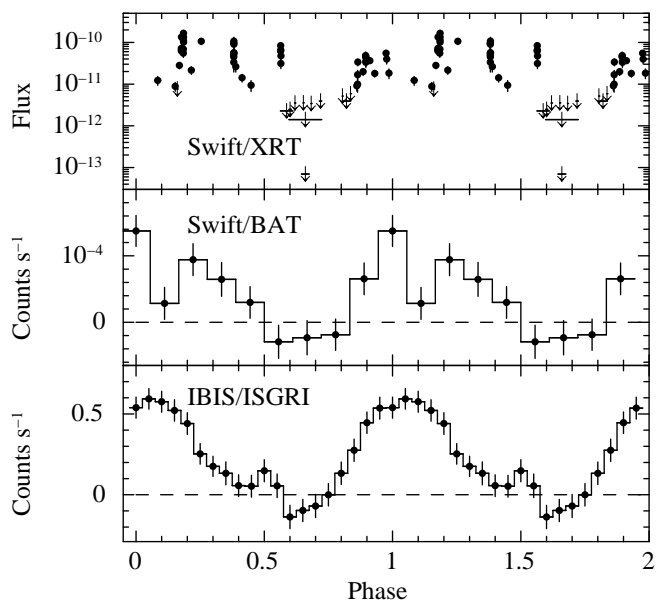}
\end{center}
\caption{Light curves of \src\ folded at the same $P_{\rm orb}$ and
$T_{\rm epoch}$ of Fig. \ref{igr17354:fig:xrtlcv_phase}. From top to bottom:  
\sw/XRT (0.3--10 keV; only the deepest XRT and \xmm\ limits are reported; 
details  in Fig.~\ref{igr17354:fig:xrtlcv_phase}), 
\sw/BAT (14--20 keV),
 and IBIS/ISGRI (17--50 keV). }
\label{igr17354:fig:folded}
\end{figure}

 	 \section{Discussion \label{igr17354:discuss} }

We performed a monitoring of the FOV around \src, which is an \inte\ transient 
positionally associated with the gamma-ray transient AGL~J1734$-$3310. 
Our new XRT data represent the first soft X-ray monitoring of more than one orbital period 
of \src, allowing us to highlight conspicuous similarities between the folded light curves 
of the XRT source and \src. Consequently, they provide a definitive identification 
of the soft X--ray counterpart of \src. 
In contrast, only a positional association, albeit one supported by 
the consistency of high and low X-ray flux with the hard X-ray flux at the same orbital phase, 
was available before. 

Based on its hard X--ray behavior, \src\ is considered a candidate SFXT by  
\citet{Sguera2011:17354}. They show that it is a weak persistent hard X-ray source
(average out-of-outburst flux of 1.1 mCrab) with rare, fast (hr to a few days, with a peak at
$\approx100$ mCrab) flares with a dynamic range $>200$. 
They also quote a soft X-ray dynamic range, which is 
based on the 2008 and 2009 XRT observations only, in excess of 300.  

The X-ray flaring activity of \src\ involves
intensity increases by a factor of $\sim 10 - 300$ on time scales
of hundreds to thousands of seconds. These increases are hardly reconciliable
with a system accreting from a disc.
In fact, an accretion disc would smooth out the accretion rate
variations produced by wind inhomogeneities with a time scale
shorter than the viscous time scale of the accretion disc.
In contrast, the observed X-ray variability, together with the
broadband X-ray spectrum and the measured orbital period \citep{Dai2011:period_17354}
is consistent with a neutron star fed by a strong stellar wind
(see, e.g., \citealt{Negueruela10}).

Our sensitive monitoring of this source in the soft X--ray  allowed us to observe
the presence of a dip at phases $0.6 \lesssim \phi \lesssim 0.85$, 
which correspond to a duration of $\Delta t_{\rm XRT} \lesssim 2.1$ d. 
This dramatic decrease in soft X-ray flux was seen  
during the \sw/XRT 2008 observation and twice in the 2012 campaign.   
It is also independently reported by a different soft X--ray
instrument in the \xmm\ 2011 observation
\citep[][]{Bozzo2012:HMXBs}.
In Fig.~\ref{igr17354:fig:folded}b and c we show the 
BAT and IBIS/ISGRI folded light curves.
The dip observed in the BAT data nicely corresponds to what is
observed in the XRT folded light curve. When fitted with a sinusoidal
function, it has a centroid at $\phi\approx0.70$ and a width 
(corresponding to the portion of the sinusoidal function with counts below zero)
spanning from 0.57 and 0.81 in phase. 
We modeled the IBIS/ISGRI light curve, which has a much larger count statistics
than the BAT one, with two sinusoidal functions and 
obtained a minimum at $\phi \approx 0.66$ and a width spanning from 0.56 and
0.75 in phase.  
The soft X-ray dynamic range from  XRT monitoring is $\sim 18$
outside of the dip and $\sim2400$ considering the deep \emph{XMM-Newton}
upper-limit.

\begin{figure}
\begin{center}
\includegraphics[bb=126 312 480 558,clip,width=9cm]{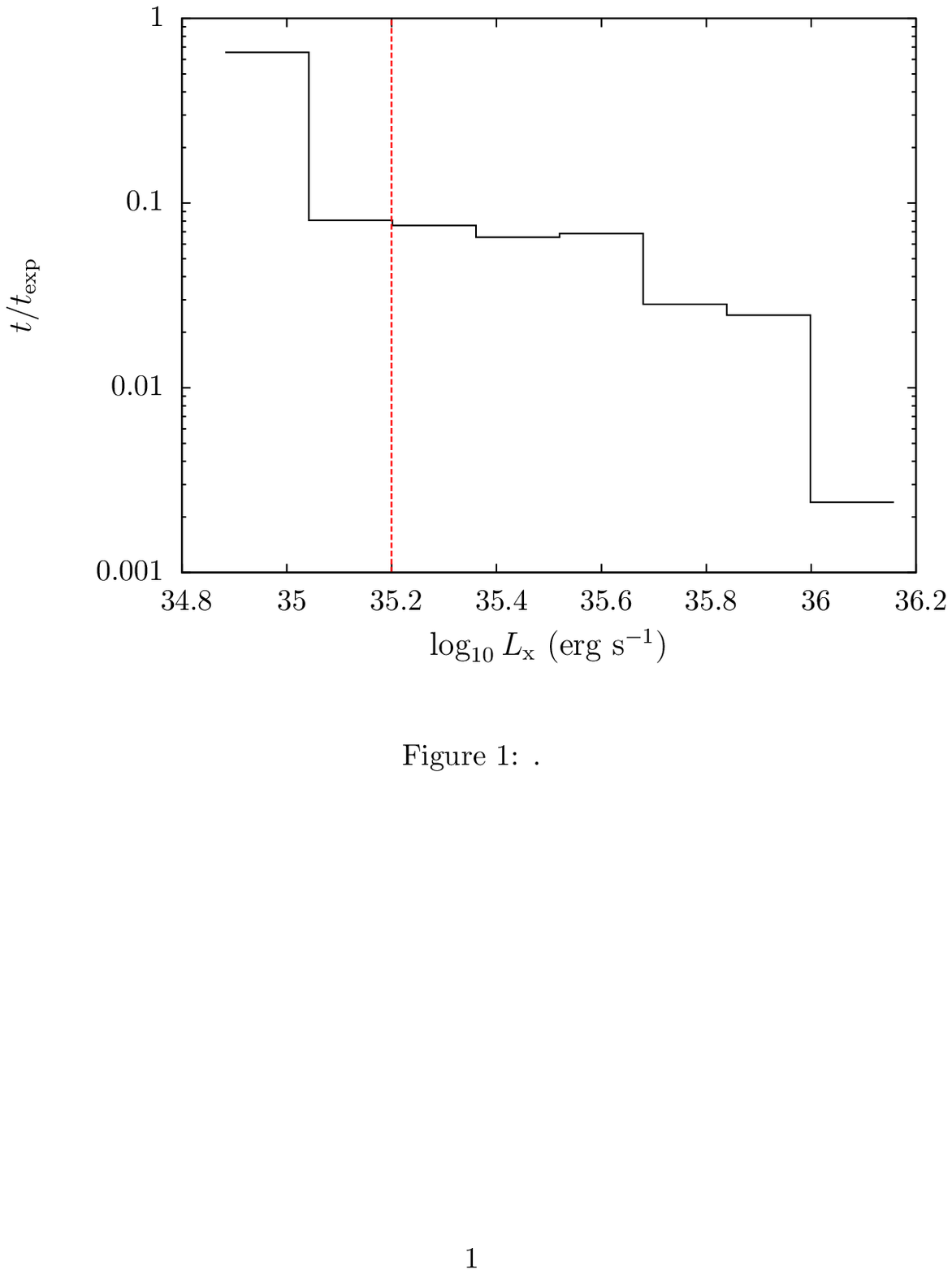}
\end{center}
\caption{Histogram of probability distribution of luminosity of \src, observed with XRT.
The vertical dashed line corresponds to the luminosity threshold below which the 
probability distribution of luminosity is not well reproduced 
due to the sensitivity of XRT.}
\label{igr17354:fig:histoobs}
\end{figure}

Fig. \ref{igr17354:fig:histoobs} shows the probability distribution histogram
of luminosity of \src, which was obtained with the XRT observations 
in which the source is significantly detected.
Because of the sensitivity of XRT, luminosities below 
$L_{\rm x}=1.6 \times 10^{35}$~erg~s$^{-1}$ 
have been obtained with large integration times.
Therefore, the luminosity distribution on
the left side of the vertical dashed line of Fig. \ref{igr17354:fig:histoobs}
does not reproduce the real luminosity distribution of \src.

\src\ looks intrinsically less variable in the soft X--ray than other SFXTs  
monitored with XRT along their orbital periods. 
In the case of the confirmed SFXT IGR~J18483--0311, 
which many consider to be an intermediate SFXT, 
the light curve shows a large modulation with the orbital phase 
\citep[][fig.~1]{Romano2010:sfxts_18483}, which 
can be interpreted as wind accretion along a highly eccentric orbit. 
The dynamic range of this source, which was calculated excluding the dip in the light curve, 
exceeds 580 (and 1200 including the dip data). 
Furthermore, variability is observed on short time scales, 
superimposed on the long-term orbital modulation of IGR~J18483--0311.
It has variations by a factor of a few 
in count rate occurring in $\sim 1$ hr, which can be naturally explained by clumps 
in the accreting wind.  
In the case of IGR~J16418--4532 \cite[][fig.~4]{Romano2012:sfxts_16418},  
in contrast, the X-ray light curve does not show a very strong orbital 
modulation, suggesting that the system has moderate eccentricity. 
The observed dynamical range of this source is at least 370, 
(1400 considering the points within the observed eclipse) and 
also places  IGR~J16418--4532 among the  intermediate SFXTs. 

Similar to IGR~J16418--4532, \src\ has a light curve with very little orbital 
modulation, and the short time-scale flares we observe superimposed on it
are comparatively of a lower dynamical range. 
Therefore, our data indicate that this source is a weak, almost persistent source 
in the soft X-rays (apart from the dip), 
which has not shown any remarkable activity during the 2012 XRT monitoring 
or in the previous observations. 

The X-ray variability of IGR J17354--3255 is also very similar
to that shown by Vela X--1, a persistent wind-fed HMXB composed of a
B0.5Ib star and a pulsar with an orbital period of 8.96 d
(similar to the orbital period of \src)
and an eccentricity of $\sim 0.09$ (e.g., \citealt{Quaintrell2003}).
Vela X--1 shows an X-ray variability with a dynamic range of 20--30
on time scales of a few hours and an average luminosity of
$\sim 4\times 10^{36}$ erg s$^{-1}$ (\citealt{Fuerst2010}; \citealt{Kreykenbohm2008}).
The lower luminosity of \src\ compared with Vela X--1
can be explained by reasonable values for the stellar wind parameters
of the donor star.

On the basis of the above considerations, it is reasonable to assume
that \src\ is a wind-fed system. In this scenario, we discuss three 
different hypotheses for the origin of the dip of \src:
\begin{enumerate}
\item In the framework of wind accretion along a highly eccentric orbit,
      the dip is due to the apastron passage of the compact object,
      where the faster and less dense wind reduces the amount of accreted material
      and consequently the X-ray luminosity.
\item The dip is caused by the onset of gating mechanisms
      at apastron, which become effective owing to the lower accretion rate (see, e.g., \citealt{Bozzo2008}).
\item The dip is produced by an eclipse.
\end{enumerate}
We discuss these three different scenarios by comparing
the X-ray luminosities observed by \emph{Swift}/XRT at different orbital phases
with those calculated with a model
based on the Bondi-Hoyle-Lyttleton accretion theory 
(\citealt{Bondi1952}; \citealt{Bondi1944}; \citealt{Hoyle1939}, BHL hereafter).
The BHL accretion theory is usually applied to X-ray binaries
where the donor star produces a fast and dense stellar wind that is assumed to 
be homogeneous.

Since the spectral type of the donor star of \src\ is unknown,
we considered both the accretion from a spherically symmetric wind
(if the donor star is a supergiant or a giant/main sequence star without circumstellar disc)
and the accretion from a circumstellar disc (if the donor star is a giant/main sequence).
The equations that we used 
to compute the luminosity light curves are described in the following paragraphs.
The results are discussed in Sects. \ref{sect. orbital modulation}  and \ref{sect. eclipse or c.i.}.

\subsection*{OB supergiant}
The winds of OB supergiants are spherically symmetric 
with good approximation (e.g., \citealt{Kudritzki00}) with a velocity law:
\begin{equation} \label{eq. beta-velocity law}
v(r) \simeq v_\infty \left ( 1 - \frac{R_{\rm d}}{r} \right )^\beta ,
\end{equation}
called $\beta$-velocity law (\citealt*{Castor1975}), where
$v_\infty$ is the terminal velocity, $\beta$ determines how
steeply the wind velocity reaches $v_\infty$ ($0.5 \lesssim \beta \lesssim 1.5$),
and $r$ is the distance from the center of the supergiant star.

The density distribution around the donor star is given by the continuity equation:
\begin{equation} \label{eq. continuity}
\rho(r) = \frac{\dot{M}}{4 \pi r^2 v(r)} ,
\end{equation}
where $\dot{M}$ is the mass loss rate.
We applied  the definition of accretion radius $R_{\rm acc} $ of \citet{Bondi1952}
to find the mass accretion rate $\dot{M}_{\rm acc}$:
\begin{equation} \label{eq. Macc}
\dot{M}_{\rm acc} = \rho(r) v_{\rm rel}(r) \pi R_{\rm acc}^2(r) = \rho(r) v_{\rm rel}(r) \pi \left[ \frac{2GM_{\rm ns}}{v_{\rm rel}^2(r)} \right ]^2 ,
\end{equation}
where $v_{rel}(r)$ is the relative velocity between the neutron star and the wind:
\begin{equation} \label{eq vel}
v_{\rm rel}(r) = \left \{ [v(r) - v_{\rm r}(r)]^2 + v_\phi^2(r) \right \}^{1/2} ,
\end{equation}
where $v_{\rm r}$ and $v_\phi$ are the radial and tangential components of the orbital velocity
and $v$ is the wind velocity (Equation \ref{eq. beta-velocity law}). We thus obtain
the X-ray luminosity produced by the accretion 
\begin{equation} \label{eq Lx sgxb}
L_x \approx \frac{G M_{\rm ns}}{R_{\rm ns}} \dot{M}_{\rm acc} = \frac{(G M_{\rm ns})^3}{R_{\rm ns}} \frac{4 \pi \rho(r)}{v_{\rm rel}^3(r)}.
\end{equation}

\subsection*{OB giants/main sequence}
The rapidly rotating Oe and Be main sequence/giant stars can be surrounded 
by a circumstellar envelope of gas confined along the equatorial plane.
Most OBe/X-ray binaries are transient in X-rays, with eccentricities $e \gtrsim 0.3$.
They can show periodic or quasi-periodic outbursts (called \emph{type I}),
which cover a small fraction of the orbital period 
and peaked at the periastron passage of the neutron star,
or they can show giant outbursts (called \emph{type II}),
which last for a large fraction of the orbit
and in some cases for several orbital periods.
These giant outbursts have peak luminosities 
of $L_{\rm x} \gtrsim 10^{37}$ erg s$^{-1}$, 
which are larger than those observed during type I outbursts.
Type I and II outbursts are believed to be due to the interaction between the neutron star
and the circumstellar disc of the donor star.
Persistent OBe/X-ray binaries do not display large outbursts and their X-ray luminosities
are $L_{\rm x} \lesssim 10^{35}$ erg s$^{-1}$ (see, e.g., \citealt{Reig2011}).

In the framework of the wind model developed by \citet{Waters1989} 
to reproduce the X-ray luminosities of wind-fed neutron stars
in eccentric orbits around the Oe-Be stars,
we assumed a density distribution in the circumstellar disc
\begin{equation} \label{eq. density disc}
\rho_{\rm disc}(r) = \rho_0 \left( \frac{r}{R_{\rm d}} \right )^{-n} ,
\end{equation}
where $\rho_0=10^{-11}$ g cm$^{-3}$, and $2.1<n<3.8$ \citep{Waters1988}.
The radial wind velocity component is
\begin{equation} \label{eq. radial wind velocity}
v_{\rm r,w}(r) = v_0 \left ( \frac{r}{R_{\rm d}} \right )^{n-2} ,
\end{equation}
where $v_0$ ranges between $2$ and $20$ km s$^{-1}$.
The rotational wind velocity component is
\begin{equation} \label{eq rotational wind velocity}
v_{\rm rot,w} (r) = v_{\rm rot,0} \left ( \frac{r}{R_{\rm d}} \right )^{-\alpha} ,
\end{equation}
where $\alpha=0.5$ for Keplerian rotation or $\alpha=1$ if the angular momentum 
of the outflowing matter is conserved.
The relative velocity between the neutron star and the wind is
\begin{equation} \label{eq. rel velocity disc}
v_{\rm rel,disc}(r) = \left \{ [v_{\rm r,w}(r) - v_{\rm r}(r)]^2 + [v_{\rm rot,w}(r) - v_\phi(r)]^2 \right \}^{1/2} .
\end{equation}
If the circumstellar disc is not formed (or outside of it)
the wind properties are the same of those found for OB supergiants,
with lower mass loss rates and terminal velocities
($10^{-10} \lesssim \dot{M} \lesssim 10^{-8}$ $M_\odot$ yr$^{-1}$;
$600 \lesssim v_\infty \lesssim 1800$ km s$^{-1}$; see, e.g., \citealt{Waters1988}).

\begin{figure*}
\begin{center}
\hspace{-1truecm}
\includegraphics[bb=172 325 416 535,clip,width=8cm]{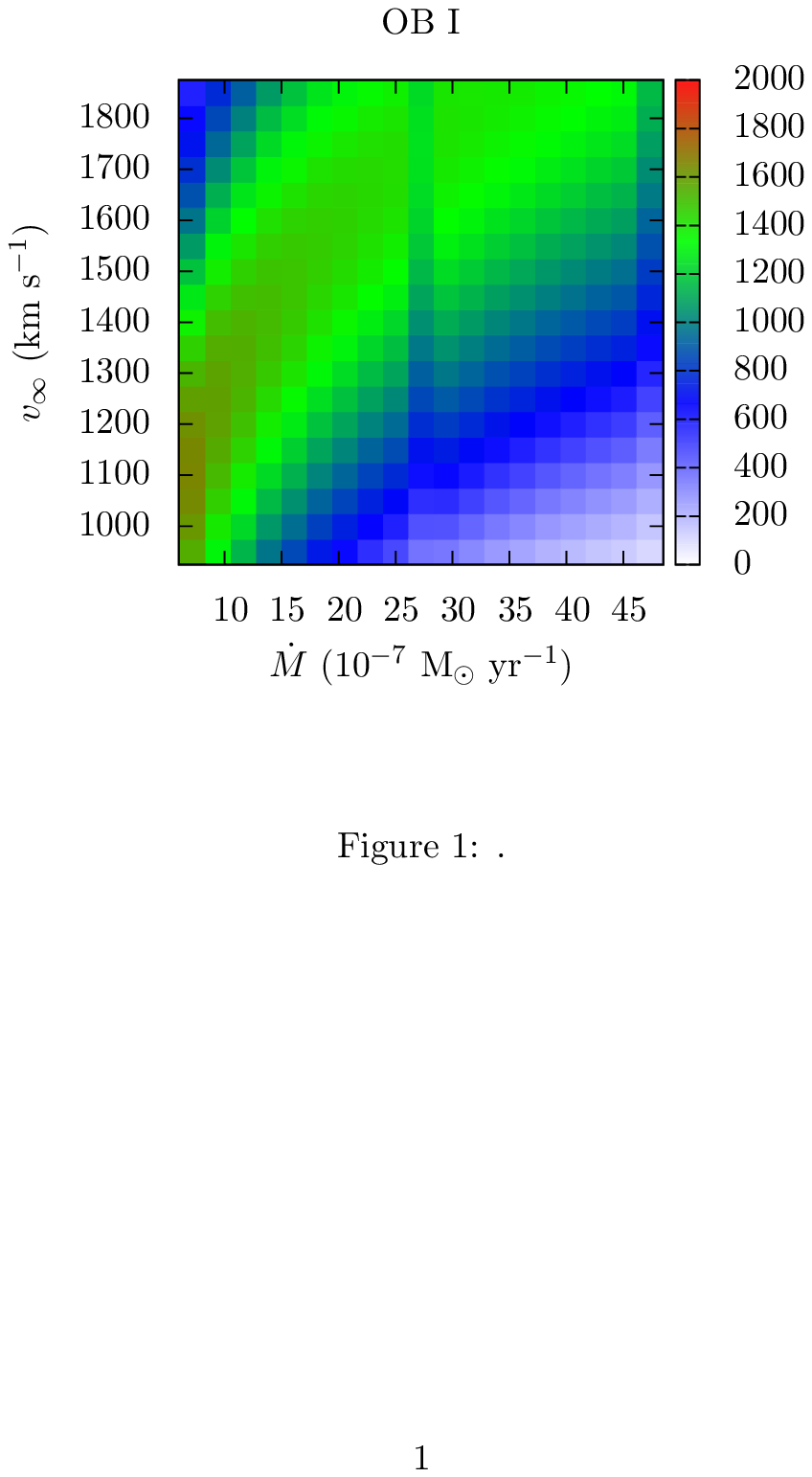}
\includegraphics[bb=172 325 416 535,clip,width=8cm]{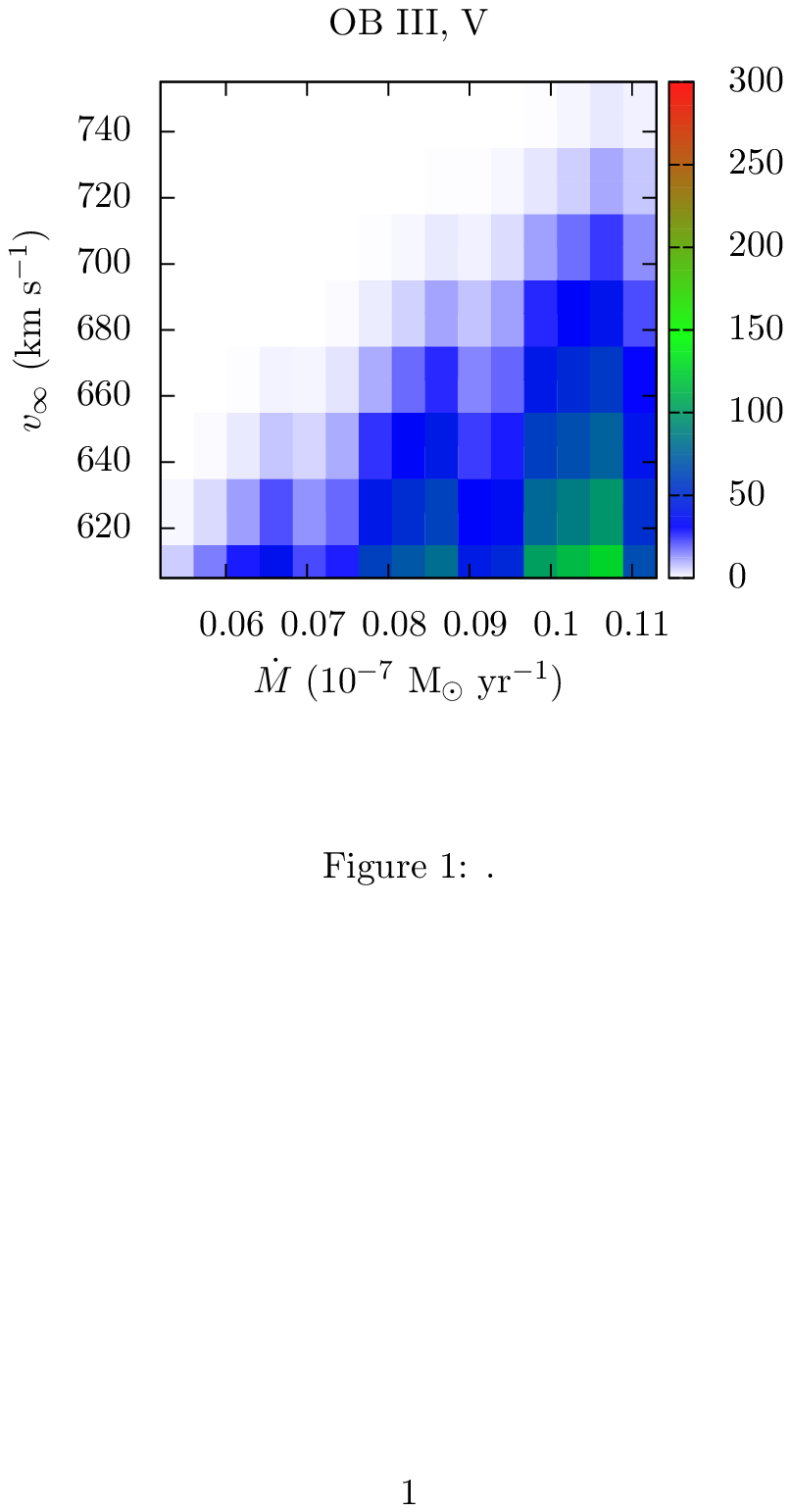}
\end{center}
\caption{2D histograms obtained with a BHL model, 
assuming a neutron star accreting a spherically symmetric wind produced 
by a supergiant (left panel) or a main sequence/giant star (right panel). 
Histograms show the solutions that reproduce the out-of-dip
luminosities. Different colors refer to different numbers of occurrences.}
\label{igr17354:fig:histo2d}
\end{figure*}

         \subsection{Orbital modulation} \label{sect. orbital modulation} 
First we considered the possibility that the \emph{XMM-Newton} dip is due to
the luminosity modulation produced by a neutron star moving in a highly eccentric orbit. 
For this calculation we did not take into account the 
duration of the dip observed by XRT,
which can only be considered as an upper limit.

Since the broadband spectrum is typical of accreting neutron stars
in HMXBs (\citealt{Sguera2011:17354}; \citealt{Dai2011:period_17354}),
we focused our attention on the case of a neutron star
with mass $M_{\rm ns}=1.4$ $M_\odot$ and radius $R_{\rm ns}=10$ km
accreting the wind material of an OB star.
We calculated the X-ray luminosities as a function of the orbital phase,
assuming a neutron star accreting material in a circumstellar disc 
or in a spherically symmetric wind and a 
supergiant, giant, main-sequence companion star.

We first computed $3\times 10^{6}$ light curves\footnote{The numbers of light curves
computed in the three cases of accretion from a spherically symmetric wind
with (1) supergiant, (2) giant or main sequence donor star, 
(3) accretion from a circumstellar disc,
depend on the number of steps used for the orbital and wind parameters
and their allowed ranges.}
by varying the masses and radii of OB supergiants (using the values reported 
in \citealt{Martins2005} and \citealt{Searle2008})
and for different values of mass loss rate $5\times 10^{-7} <\dot{M}<5\times 10^{-6}$ $M_\odot$ yr$^{-1}$,
terminal wind velocity $900<v_\infty<1900$ km s$^{-1}$, and
eccentricities $0<e<0.8$.
We set $\beta=1$ because we ascertain that variations of this parameter
do not produce appreciable variations in the X-ray luminosity.

Then, we considered the case of OB giant/main sequence stars with circumstellar disc.
We calculated $2.5\times 10^{6}$ light curves by varying the masses and radii of OB\,V
(using the mass-radius relation of \citealt{Demircan1991}) and
OB\,III (using the Catalogue of Apparent Diameters and Absolute Radii of Stars 
[CADARS] of \citealt{Pasinetti-Fracassini2001}; \citealt{Hohle2010};
\citealt{Martins2005}), and for different wind parameters and eccentricities:
$2.1 \lesssim n \lesssim 3.8$, $2 \lesssim v_0 \lesssim 22$ km s$^{-1}$,
$150 \lesssim v_{\rm rot,0} \lesssim 310$ km s$^{-1}$,
$\alpha=0.5$ or $1$, $0<e<0.8$.

In the scenario of OB giant/main sequence stars without the circumstellar discs,
we calculated $1.5\times 10^{6}$ light curves assuming different spectral types for the donor star,
mass loss rates $10^{-10} \lesssim \dot{M} \lesssim 10^{-8}$ $M_\odot$ yr$^{-1}$,
terminal velocities $600 \lesssim v_\infty \lesssim 1800$ km s$^{-1}$,
eccentricities $0<e<0.8$, and $\beta=1$.

We computed the X-ray luminosities using Equation (\ref{eq Lx sgxb}),
with $v_{\rm rel}(r)$ and $\rho(r)$ given by Equations
(\ref{eq vel}) and (\ref{eq. continuity})
in the spherically symmetric wind scenario
and by Equations (\ref{eq. rel velocity disc}) and (\ref{eq. density disc}) 
if the donor star is an OB main sequence/giant and a circumstellar disc is present.

In all cases, no combination of the orbital and wind parameters 
can reproduce the observed out-of-dip ($0 < \phi < 0.6$ and $0.85 < \phi < 1$)
luminosities ranging from $\sim 6 \times 10^{34}$ erg s$^{-1}$
to $\sim 3 \times 10^{36}$ erg s$^{-1}$ ($d=8.5$ kpc)
and luminosities below $4 \times 10^{32}$ erg s$^{-1}$
at $\phi \approx 0.65$ (corresponding to the \emph{XMM-Newton} observation).
In fact, the observed dynamic range (considering the \emph{XMM-Newton} upper limit) 
requires highly eccentric orbits. However, when $e>0$ the neutron star spends
a large fraction of time in the apastron region (Kepler's second law).
Therefore, all calculated light curves reproducing the \emph{XMM-Newton} upper limit at $\phi \approx 0.65$
show luminosities lower than $\approx 10^{34}$ erg s$^{-1}$ during orbital phases
longer than the dip observed by XRT.

In conclusion, it is unlikely that the upper limit observed by \emph{XMM-Newton}
is due to the apastron passage of the neutron star.

\begin{figure}
\begin{center}
\hspace{-1truecm}
\includegraphics[bb=110 311 474 554,clip,height=5.5cm]{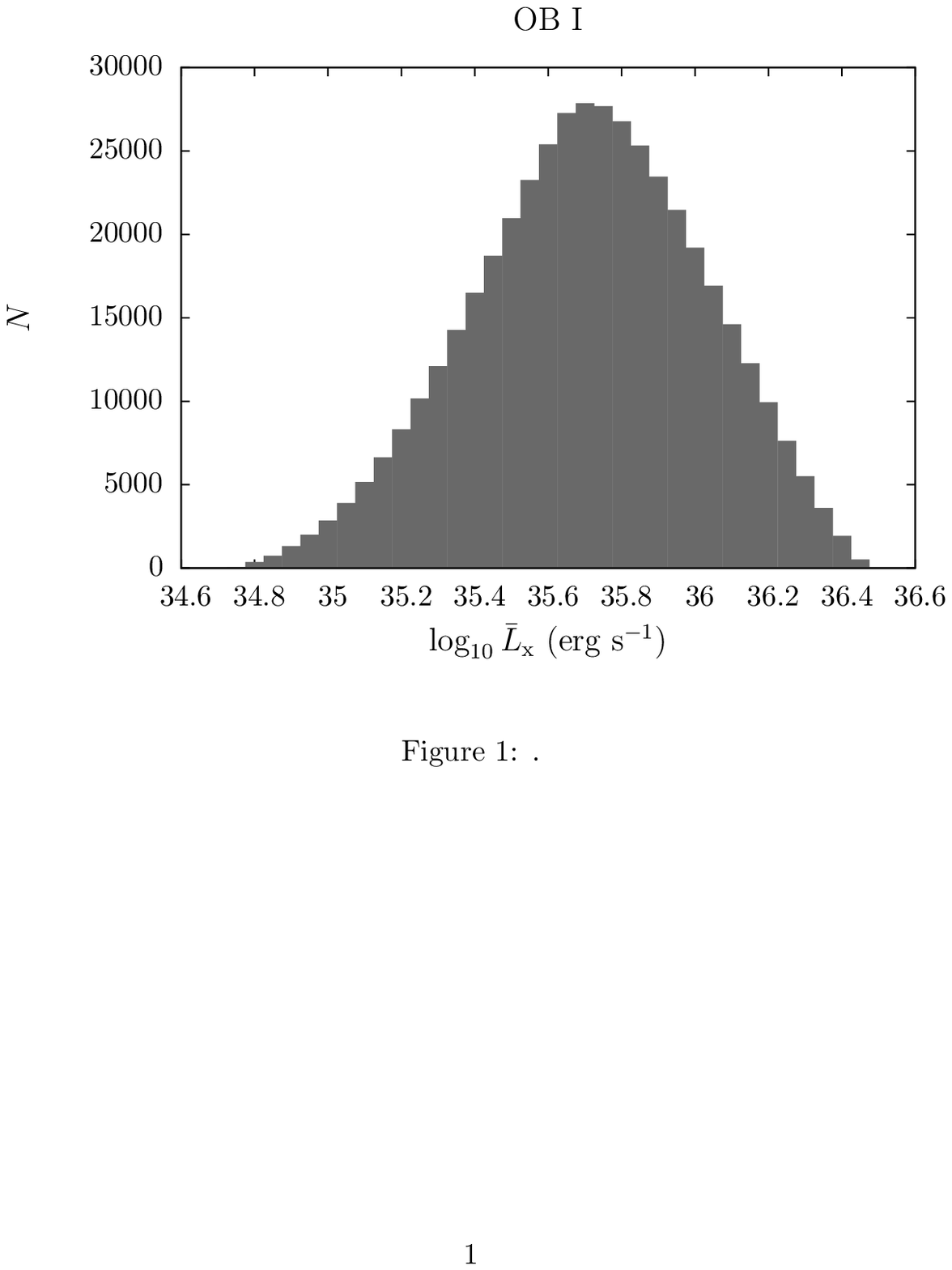}
\includegraphics[bb=125 311 474 554,clip,height=5.5cm]{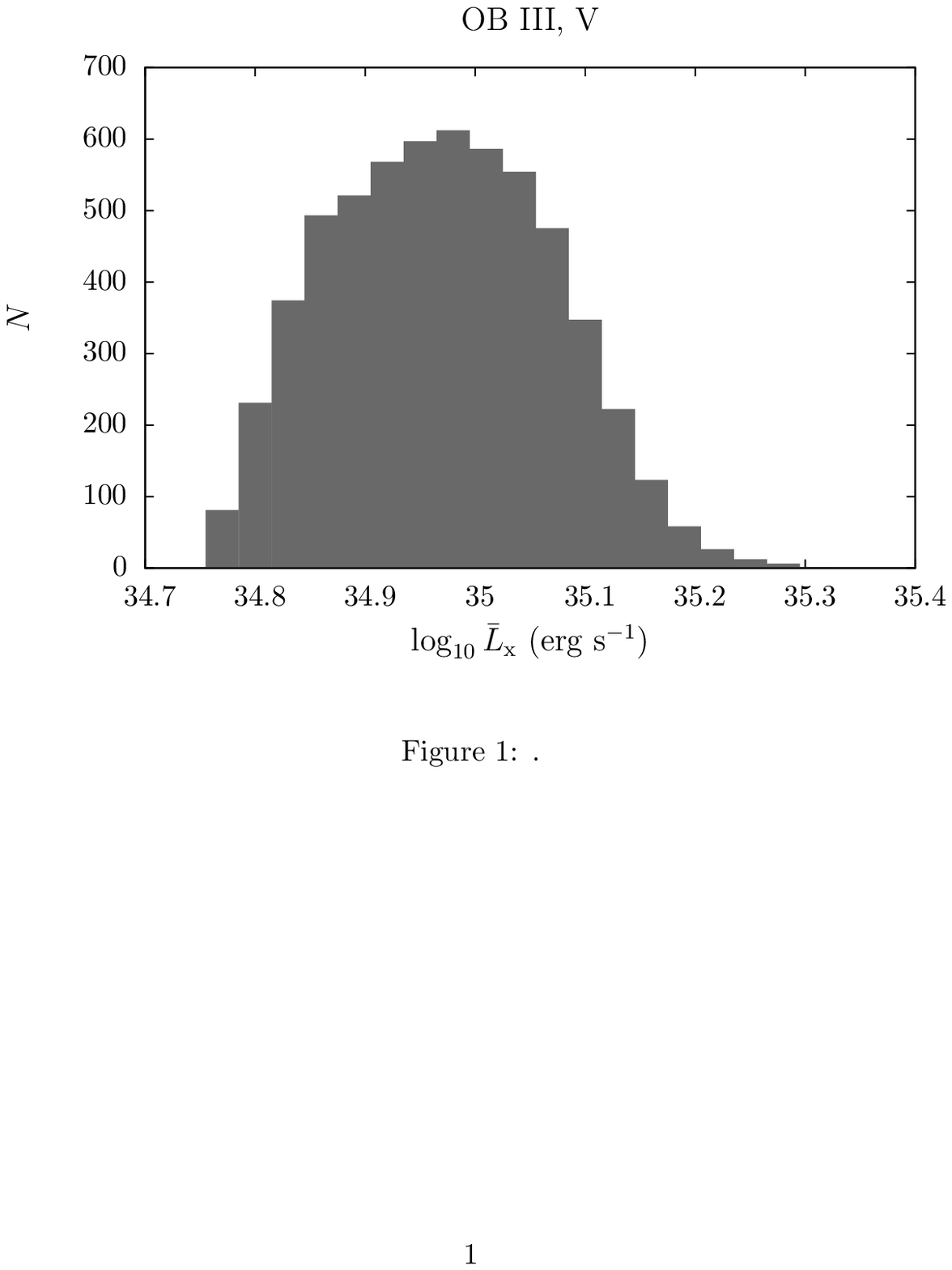}
\end{center}
\caption{Histograms of the mean luminosity (averaged over the orbit) obtained with the BHL
model, assuming the parameters giving the solutions of Fig. \ref{igr17354:fig:histo2d},
$0<e<0.64$ for an OB supergiant, $0<e<0.43$ for an OB main sequence/giant star.}
\label{igr17354:fig:histo1d}
\end{figure}

         \subsection{Eclipse and gating mechanisms} \label{sect. eclipse or c.i.} 

Similarly to Sect. \ref{sect. orbital modulation},
we compared the observed X-ray luminosities with those obtained 
with a BHL model to determine the eccentricities of the system, 
the wind properties, and the nature of the donor star that reproduces the observed
out-of-dip X-ray luminosities,
which range from $\sim 6 \times 10^{34}$ erg s$^{-1}$
to $\sim 3 \times 10^{36}$ erg s$^{-1}$ ($d=8.5$ kpc).

Since the observational properties of the dip (duration and luminosity)
in both the eclipse and gated mechanisms cases
do not depend solely on the wind and orbital parameters\footnote{In the eclipse scenario, 
the duration of the eclipse depends on the
orbital separation, the radius of the donor star, and the 
inclination of the orbital angular momentum vector with respect to the 
line of sight to the Earth. In the gated mechanisms scenario,
the dip can be produced in different epochs at a given orbital phase
if $e > 0$ and for particular values of the 
spin period and magnetic field strength of the neutron star.},
we treated these two scenarios as a single one.

The 2D histograms of Fig. \ref{igr17354:fig:histo2d} show the 
solutions that reproduce the observed X-ray luminosities
obtained assuming
the accretion from a spherically symmetric wind produced by a supergiant (left panel)
or a giant/main sequence star (right panel).
For each solution we also plotted the mean luminosity $\bar{L}_x$ (averaged over the orbit)
in the histograms of Fig. \ref{igr17354:fig:histo1d}. These histograms show that
the X-ray luminosity produced by the accretion from a supergiant wind
is on average higher than the X-ray luminosity produced 
by the accretion of the wind from a giant or a main sequence OB star.
The allowed eccentricities (obtained by comparing the observed X-ray luminosities
with those calculated with the BHL model)
are $0 \lesssim e \lesssim 0.64$ in the supergiant scenario 
and $0 \lesssim e \lesssim 0.43$ in the giant/main sequence case.

We also considered the case of a neutron star embedded in the wind of the 
circumstellar disc of the donor star.
We found that the observed X-ray luminosities can be reproduced
assuming a main sequence donor star with mass $M_{\rm d}=8$ $M_\odot$,
$R_{\rm d}=4$ $R_\odot$, $v_0=20$ km s$^{-1}$,
$n=3.5$, $150<v_{\rm rot}<300$ km s$^{-1}$, $\alpha=0.5$ or $1$, $0<e<0.02$.
The X-ray luminosities obtained using these parameters are 
about $3 \times 10^{36}$ erg s$^{-1}$.
Assuming distances larger than $8.5$ kpc, more solutions are possible.
Nonetheless, since the truncation of the circumstellar disc produced
by the neutron star (\citealt{Reig97}; \citealt{Negueruela01}; \citealt{Okazaki01})
is expected to be more efficient in systems with low eccentricities and narrow
orbits (see \citealt{Reig2011} and references therein), it is unlikely that 
the neutron star of \src\ accretes the material of the circumstellar disc.

         \section{Summary} \label{sect. summary} 

We reported on \emph{Swift}/XRT observations of the candidate SFXT \src,
which provided for the first time a definitive identification
of its soft X-ray counterpart.
They also allowed us to observe the presence of a dip in the XRT 
light curve of \src\ folded at the orbital period.
Apart from the dip, the low dynamic range observed with \emph{Swift}/XRT
indicates that \src\ is an almost persistent source in the soft X-rays.

We investigated the origin of the dip by comparing the XRT folded light curve
with those calculated with models based on the BHL accretion theory.
We assumed both spherical and nonspherical symmetry of the outflow from the donor star.
We found that the dip cannot be explained with a luminosity modulation produced
by a neutron star in a highly eccentric orbit
and showed that an eclipse or the onset of a gated mechanism can explain the dip.
We also determined the eccentricities of the system, the wind properties,
and the nature of the donor star that reproduces the observed out-of-dip luminosities.

\begin{acknowledgements}
We thank the anomymous referee  
for constructive comments which helped to improve the paper. 
We thank the {\it Swift} team duty scientists and science planners. 
We also thank the remainder of the {\it Swift} XRT and BAT teams, S.D.\ Barthelmy, 
J.A.\ Nousek, and D.N.\ Burrows in particular, for their invaluable help and support 
of the SFXT project as a whole. 
We thank P.A. Evans and C.\ Ferrigno for helpful discussions. 
We acknowledge financial contribution from the contract ASI-INAF I/004/11/0.
This work made use of the results of the Swift/BAT hard X-ray transient monitor:
http://swift.gsfc.nasa.gov/docs/swift/results/transients/
\end{acknowledgements}

\bibliographystyle{aa} 
\bibliography{igrj17354}

\end{document}